\documentclass[
]{ceurart}

\usepackage{listings}
\usepackage{hyperref}
\usepackage{listings}
\usepackage{xcolor}
\begin{document}

\copyrightyear{2022}
\copyrightclause{Copyright for this paper by its authors.
  Use permitted under Creative Commons License Attribution 4.0
  International (CC BY 4.0).}

\conference{BPM 2023 Demos \& Resources Forum, September 11--15 2023, Utrecht (NL)}

\title{A Collection of Simulated Event Logs for Fairness Assessment in Process Mining}


\author{Timo Pohl}[%
email=timo.pohl@rwth-aachen.de
]

\author[1]{Alessandro Berti}[%
orcid=0000-0001-8515-3089,
email=a.berti@pads.rwth-aachen.de
]
\cormark[1]

\author{Mahnaz Sadat Qafari}[
orcid=0000-0001-5387-2580,
email=sadatghafari@gmail.com
]

\author[1]{Wil M.P. van der Aalst}[%
orcid=0000-0002-0955-6940,
email=wvdaalst@pads.rwth-aachen.de
]

\address[1]{Process and Data Science Chair, RWTH Aachen University, Aachen, Germany}

\cortext[1]{Corresponding author.}

\begin{abstract}
The analysis of fairness in process mining is a significant aspect of data-driven decision-making, yet the advancement in this field is constrained due to the scarcity of event data that incorporates fairness considerations. To bridge this gap, we present a collection of simulated event logs, spanning four critical domains, which encapsulate a variety of discrimination scenarios. By simulating these event logs with CPN Tools, we ensure data with known ground truth, thereby offering a robust foundation for fairness analysis. These logs are made freely available under the CC-BY-4.0 license and adhere to the XES standard, thereby assuring broad compatibility with various process mining tools. This initiative aims to empower researchers with the requisite resources to test and develop fairness techniques within process mining, ultimately contributing to the pursuit of equitable, data-driven decision-making processes.
\end{abstract}

\begin{keywords}
  Fairness in Process Mining \sep
  Simulated Event Logs \sep
  Discrimination Analysis \sep
  Data-Driven Decision Making
\end{keywords}

\maketitle

\section{Introduction}
\label{sec:introduction}

The increasing integration of data science and machine learning into decision-making underscores the need for fairness \cite{barocas2017fairness,friedler2019comparative}. Despite widespread use, fields like process mining lack robust fairness measures due to a limited number of publicly available event logs \cite{qafari2019fairness,pohl2022discrimination}. This study seeks to bridge this gap, providing simulated event logs via CPN Tools to aid the evolution of fairness techniques in process mining.
The subsequent sections of this paper are organized as follows: Section \ref{sec:description} provides a detailed description of the event logs and their respective attributes. Section \ref{sec:preliminaryAnalysis} offers preliminary insights and analyses derived from these event logs. Section \ref{sec:downloading} outlines the procedure for downloading and utilizing this resource, supplemented with an illustrative example using the process mining library \emph{pm4py}. The paper culminates in Section \ref{sec:conclusion}, wherein we draw conclusions and offer a summation of our work.

\begin{table}[!t]
\caption{Basic statistics of the provided collection of event logs.}
\resizebox{\textwidth}{!}{
\begin{tabular}{lrrrrr}
\toprule
{\bf Event Log} & {\bf Number of Events} & {\bf Number of Cases} & {\bf Number of Variants} & {\bf Number of Activities} \\
\midrule
hiring\_log\_high-xes.gz & 63869 & 10000 & 386 & 12 \\
hiring\_log\_low-xes.gz & 72094 & 10000 & 296 & 12 \\
hiring\_log\_medium-xes.gz & 69054 & 10000 & 382 & 12 \\
\hline
hospital\_log\_high-xes.gz & 69528 & 10000 & 80 & 10 \\
hospital\_log\_low-xes.gz & 70037 & 10000 & 106 & 10 \\
hospital\_log\_medium-xes.gz & 70124 & 10000 & 77 & 10 \\
\hline
lending\_log\_high-xes.gz & 58822 & 10000 & 41 & 12 \\
lending\_log\_low-xes.gz & 60746 & 10000 & 31 & 12 \\
lending\_log\_medium-xes.gz & 58668 & 10000 & 33 & 12 \\
\hline
renting\_log\_high-xes.gz & 89972 & 10000 & 496 & 16 \\
renting\_log\_low-xes.gz & 96440 & 10000 & 508 & 16 \\
renting\_log\_medium-xes.gz & 105555 & 10000 & 610 & 16 \\
\bottomrule
\end{tabular}
}
\label{tbl:eventLogs}
\end{table}

\begin{table}[!b]
\caption{Sensitive attributes in the provided collection of event logs.}
\resizebox{\textwidth}{!}{
\begin{tabular}{|l|c|c|c|c|c|c|c|c|}
\hline
\textbf{Domain} & \textbf{Age} & \textbf{Citizenship} & \textbf{German Proficiency} & \textbf{Gender} & \textbf{Religion} & \textbf{Years of Education} & \textbf{Underlying Condition} & \textbf{Private Insurance} \\
\hline
Hiring & X & X & X & X & X & X & ~ & ~ \\
\hline
Hospital & X & X & X & X & ~ & ~ & X & X \\
\hline
Lending & X & X & X & X & ~ & X & ~ & ~ \\
\hline
Renting & X & X & X & X & X & X & ~ & ~ \\
\hline
\end{tabular}
}
\label{tbl:sensitiveAttributes}
\end{table}

\section{Description of the Resource}
\label{sec:description}

We present 12 distinct event logs, split evenly across four domains: hiring, healthcare, lending, and renting. Each log, embedded in the context of German society and containing 10,000 cases, has been meticulously curated and simulated using colored Petri nets\footnote{Refer to \cite{pohl2022thesis} for a comprehensive explanation of the Petri nets and their parameters.}. The logs display varying degrees of discrimination within each domain, enabling researchers to delve into real-world scenarios.
Table \ref{tbl:eventLogs} provides some basic statistics on the collection, while Table \ref{tbl:sensitiveAttributes} describes the sensitive attributes.
While the chosen attributes related to fairness are debatable, we encourage discourse to refine our understanding of fairness in process mining. Each log's attributes and process are elaborately described to aid in identifying potential sources of discrimination.

All logs comply with the eXtensible Event Stream (XES) \cite{verbeek2011xes} standard format, promoting compatibility and interoperability with diverse process mining tools, thus facilitating research across various platforms.

\subsection{Hiring Event Logs}

\noindent \emph{Process:}
The data describes a multifaceted recruitment process with diverse application pathways ranging from minimal processing to extensive multi-step procedures. The variability of these routes, largely dependent on numerous determinants, yields a spectrum of outcomes from instant rejection to successful job offers.

\noindent \emph{Attributes:}
The logs include attributes such as age, citizenship, German proficiency, gender, religion, and years of education. While these attributes may inform candidate profiles, their misuse could engender discrimination. Variables like age and education may signify experience and skills, citizenship, and German language may address job logistics, but these should not unjustly eliminate applicants. Gender and religion, unrelated to job performance, must not sway hiring. Therefore, the use of these attributes must uphold fairness, avoiding any potential bias.

\subsection{Hospital Event Logs}

\noindent \emph{Process:}
The data depicts a hospital treatment process that commences with registration at an Emergency Room or Family Department and advances through stages of examination, diagnosis, and treatment. Notably, unsuccessful treatments often entail repetitive diagnostic and treatment cycles, underscoring the iterative nature of healthcare provision.

\noindent \emph{Attributes:}
The logs incorporate patient attributes such as age, underlying condition, citizenship, German language proficiency, gender, and private insurance. These attributes, influencing the treatment process, may unveil potential discrimination. Factors like age and condition might affect case complexity and treatment path, while citizenship may highlight healthcare access disparities. German proficiency can impact provider-patient communication, thus affecting care quality. Gender could spotlight potential health disparities, while insurance status might indicate socio-economic influences on care quality or timeliness. Therefore, a comprehensive examination of these attributes vis-à-vis the treatment process could shed light on potential biases or disparities, fostering fairness in healthcare delivery.

\subsection{Lending Event Logs}

\noindent \emph{Process:}
This data illustrates the steps within a loan application process. From an initial appointment request, the process navigates various stages, including information verification and underwriting, culminating in loan approval or denial. Additional steps may be required, such as co-signer enlistment or collateral assessment. Some cases experience outright appointment denial, indicating the process's variability, reflecting applicants' differing credit situations.

\noindent \emph{Attributes:}
The logs' attributes can aid in identifying influences on outcomes and detecting discrimination. Personal characteristics ('age', 'citizen', 'German speaking', and 'gender') and socio-economic indicators ('YearsOfEducation' and 'CreditScore') can impact the process. While 'YearsOfEducation' and 'CreditScore' can validly inform creditworthiness, 'age', 'citizen', 'language ability', and 'gender' should not bias loan decisions, ensuring these attributes are used responsibly fosters equitable loan processes.

\subsection{Renting Event Logs}

\noindent \emph{Process:}
The data encapsulates a detailed rental process, starting from an application to view a property to potential contract termination. Intermediate steps include initial screening, property viewing, decision-making, and possibly, a thorough screening. If approved, a rental agreement commences, where late payments can lead to eviction, one potential process endpoint. Alternatively, tenants might voluntarily end their contracts. Note that not all applications proceed to the viewing stage.

\noindent \emph{Attributes:}
The logs contain attributes that can shed light on potential biases in the process. 'Age', 'citizen', 'German speaking', 'gender', 'religious affiliation', and 'yearsOfEducation' might influence the rental process, leading to potential discrimination. While some attributes may provide useful insights into a potential tenant's reliability, misuse could result in discrimination. Thus, fairness must be observed in utilizing these attributes to avoid potential biases and ensure equitable treatment.

\section{Preliminary Analysis}
\label{sec:preliminaryAnalysis}

Performing an initial analysis of discrimination across the four contexts - hiring, hospital, lending, and renting - and primarily focusing on the
control-flow (the sequence of activities) and time perspectives, notable disparities between the 'protected' and 'non-protected' groups (identified by the values in the cases' attributes) were observed. These discerned discriminations, specifically in the context of high discrimination, can be summarized as follows:
\begin{itemize}
\item In the hiring process, the protected group encounters higher rejection rates at the application stage, is provided fewer opportunities for interviews, and receives fewer job offers compared to the non-protected group.
\item Within the hospital logs, the protected group tends to follow more complex and longer treatment paths. They undergo 'Thorough Examination' less often and face unsuccessful treatments more frequently than the non-protected group. Additionally, it has been observed that the protected group is predominantly managed by less experienced doctors, potentially contributing to their more complex and lengthy treatment journeys.
\item In the lending logs, the protected group experiences higher rates of appointment denials and loan application rejections. They also often go through additional steps such as collateral assessment and co-signer requests, and face a higher rate of loan denials.
\item Finally, for renting, the protected group faces higher immediate rejection rates after viewing appointment applications and undergoes 'Extensive Screening' more frequently. Interestingly, they also tend to remain longer in their apartments once accepted as tenants.
\end{itemize}
By offering quantifiable instances of discrimination across varying contexts, these logs demonstrate their substantial potential as a resource for furthering the exploration of fairness in process mining, thereby aiding the scientific community in fostering more equitable and just processes.

\lstset{language=Python}
\lstset{frame=lines}
\lstset{caption={Example \emph{pm4py} code to use one of the provided event logs.}}
\lstset{label={lst:pm4pyFairness}}
\lstset{basicstyle=\scriptsize}
\begin{lstlisting}
import pm4py
log = pm4py.read_xes("hiring_log_high.xes.gz") # reads the resource
# splits the log into the protected and non-protected groups.
protected_log = log[log["case:protected"] == True]
nonprotected_log = log[log["case:protected"] == False]
# discovers different process models for the groups using the inductive miner
protected_model = pm4py.discover_process_tree_inductive(protected_log)
nonprotected_model = pm4py.discover_process_tree_inductive(nonprotected_log)
# shows the process models on the screen
pm4py.view_process_tree(protected_model, format="svg")
pm4py.view_process_tree(nonprotected_model, format="svg")
\end{lstlisting}

\section{Downloading and Using the Resource}
\label{sec:downloading}

The event logs \cite{pohl2023collection} are publicly accessible under the CC-BY-4.0 license at \url{https://doi.org/10.5281/zenodo.8059488}. These logs can be downloaded and utilized by any process mining tool supporting the XES standard. In Listing \ref{lst:pm4pyFairness}, we demonstrate how to use the Python library \emph{pm4py} to ingest the hiring (high discrimination) log, split the cases between the protected and unprotected groups, and discover different models for each group.

\section{Conclusion}
\label{sec:conclusion}

This work is a pivotal step towards embedding fairness in process mining, offering a distinct collection of simulated event logs across diverse domains. These logs fill a critical gap in data resources, which hinders progress in fairness-oriented process mining.

Our initial analysis, though revealing, is rudimentary, highlighting inherent limitations. As simulations, the logs might not encapsulate all real-world process complexities. Yet, they establish a foundation for developing and testing fairness measures.

Future plans include expanding this collection with privacy-preserving, real-world logs, and leveraging insights from this study to advance fairness-oriented process mining techniques. Our resources invite researchers to further explore and compare fairness techniques within process mining.

\bibliography{paper}

\end{document}